# Vanadium Doped Magnetic MoS$_2$ Monolayers of Improved Electrical Conductivity as Spin-Orbit Torque Layer

Krishna Rani Sahoo,[1,2,#] Manoj Talluri,[3,#] Dipak Maity,[1,#] Suman Mundlia,[1] Ashique Lal,[1] M. S. Devapriya,[4] Arabinda Haldar,[4] Chandrasekhar Murapaka,[3,*] and Tharangattu N. Narayanan,[1,*]

[1]*Materials & Interface Engineering Laboratory, Tata Institute of Fundamental Research Hyderabad, Serilingampally Mandal, Hyderabad 500046, India.*

[2]*Institute of Physics, University of Münster, Wilhelm-Klemm-Str. 10, 48149 Münster, Germany*

[3]*Department of Materials Science and Metallurgical Engineering, Indian Institute of Technology-Hyderabad, 502284, Telangana, India*

[4]*Department of Physics, Indian Institute of Technology-Hyderabad, 502284, Telangana, India*

(*Corresponding authors: tnn@tifrh.res.in (T.N.N.) and mchandrasekhar@msme.iith.ac.in (C.M.) )

**ToC**

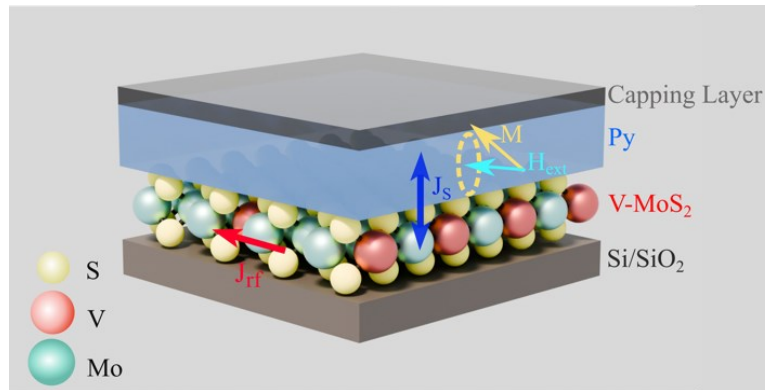

Demonstration of spin transport at the interface between permalloy (Py) and vanadium doped MoS$_2$ (V-MoS$_2$) magnetic monolayers having high electrical conductivity and spin-orbit coupling.

**Abstract:** Two-dimensional (2D) transition metal di-chalcogenide layers with high electrical conductivity and spin-orbit coupling (SOC) can find huge potential in spintronic devices. With limited success of 2D spin Hall material development, we demonstrate vanadium (V) substitutionally doped monolayer $MoS_2$ (VMS) as a potential spin Hall material having tunable electrical conductivity, SOC strength, and room temperature magnetism. Systematic enhancement in the electrical conductivity is observed with the extent of V doping, where it is enhanced from ~$3\times10^{-1}$ S/m of $MoS_2$ to ~$10^5$ S/m upon doping to the level of 9 atomic%. Ferromagnetic resonance (FMR) based spin-pumping experiments indicate the spin transport across the junction of permalloy (Py) and VMS. Spin-torque FMR measurements demonstrate the charge to spin conversion at the Py/VMS interface suggesting latter's potential as a spin-orbit torque layer in 2D spintronic devices.

**Introduction:**

Spin-orbit torque (SOT) based magnetic tunnel junctions (MTJs) have been explored extensively because of their reliable switching and quasi-infinite endurance compared to the spin transfer torque (STT) MTJs.[1–3] Typical SOT-MTJs consist of four major layers, namely the spin Hall layer - which converts the charge current to spin current through a phenomenon called the spin Hall effect (SHE), and two ferromagnetic layers - namely the free layer and the fixed layer separated by a tunneling barrier.[3,4] The spin current generated from the spin Hall layer induces torque on the free layer and changes the magnetization direction of the free layer. Tungsten, tantalum, and platinum are some of the spin Hall materials where the bulk spin Hall effect is the reason for the spin current generation. However, in 2-dimensional (2D) materials, the spin current mainly originates due to the interfacial Rashba effect rather than the bulk spin Hall effect.[5] In this process, high spin-orbit coupling (SOC) strength and high electrical conductivity are the essential criteria of spin Hall materials, which allow the efficient conversion of the charge current to spin current quantified *via* spin Hall efficiency.[6–8]

Conventional spin Hall materials, such as tungsten (W), tantalum (Ta), etc. exhibit different SOC strengths in their allotropic forms.[9,10] For example, the β phase of W exhibits higher SOC efficiency but has a higher electrical resistance than the α phase of W.[11] A similar phenomenon is seen in the case of Ta, where the β phase of Ta with high SOC strength is also observed to be resistive in nature.[8,9,12] This is due to the structural modification and increased scattering events of the conducting electrons. As a result, even though the spin Hall efficiency of such phases increases, the overall spin Hall conductivity (Spin Hall efficiency times electrical conductivity) cannot be improved. Spin Hall conductivity determines the practical adaptability of any spin Hall material in SOT-based devices. Currently, platinum (Pt) is being extensively explored as spin Hall material in SOT-based devices due to its high spin Hall conductivity because of very high electrical conductivity, even though the spin Hall efficiency is smaller compared to Ta and W.[13] However, developing low-cost materials with higher spin Hall conductivity, which combines high spin Hall efficiency and high electrical conductivity, is essential.

Layered transition metal dichalcogenides (TMDs) such as $MoS_2$, $WS_2$, $MoSe_2$, etc., in their monolayer limit, have attracted considerable interest in various fields such as electronics, optoelectronics, and spintronics due to their direct bandgap, disrupted inversion symmetry and the tunability in electronic properties along with their relatively high SOC strength.[14-16] Employing such monolayer TMDs in the spintronic devices brings in novel quantum effects (QSHE) due to the quantum confinement effects and also facilitates the miniaturization of the devices. It is reported that incorporating transition metals (V, Mn, Fe, Co, Cu etc.) into such 2D TMDs through doping or alloying is a potential method for tuning electronic, magnetic properties and SOC.[17,18] Among known TMDs, $MoS_2$ is one of the most studied semiconducting 2D TMDs, having significantly high SOC and chemical stability.[19] Multiple studies have shown the possibility of chemical vapor deposition (CVD) technique to achieve wafer-scale growth of 2D-$MoS_2$ layers.[20,21] The CVD-grown $MoS_2$ layers are known for their n-type carrier transport, believed to be due to the sulfur vacancies present, with electrical conductivity <1 S/m.[14,22] Upon doping the monolayer $MoS_2$ with V, it is shown that the charge carrier density increases, improving the electrical conductivity significantly.[22] The V atoms substitute Mo in $MoS_2$, as shown by the authors in their previous works.[18,22,23] As the V atom with a low atomic weight (atomic number) is replacing Mo with a high atomic weight (atomic number), the effective SOC strength of the layers is predicted to be modified.

Density functional theory (DFT) based calculations indicate that doping of V in $MoS_2$ (at~ 8 atomic%) can induce ferromagnetic ordering in these monolayers.[24] Gao *et al.* have proposed induced ferromagnetism with high $T_C$ in V-doped $MoS_2$ (VMS) monolayers, particularly at relatively high V concentrations.[25] Studies suggest that the ferromagnetism in VMS originates from the lattice strain, defects such as Mo and S vacancies, and, more importantly, V-V ferromagnetic interaction.[26,27] This indicates that the electrical conductivity, SOC strength, and ferromagnetic ordering of the $MoS_2$ layer can be tuned by varying V doping and one has to reach a tradeoff among all by optimizing the V content to achieve high spin Hall conductivity.

Here we, for the first time, demonstrate the development of VMS with tunable electrical conductivity and SOC strength along with the induced ferromagnetic ordering at room temperature. $MoS_2$ monolayers (MS) are grown using a conventional CVD deposition method, and an *in situ* doping approach is followed for the development of VMS during the $MoS_2$ growth. The extent of doping has been tuned from 1 atomic% to 9 atomic%, as discussed in the experimental section. To perform spin pumping studies into MS and VMS, heterostructures of magnetic permalloy (Py) thin films with MS and VMS are prepared where the Py films are deposited on MS and VMS using magnetron sputtering technique. The spin-pumping studies, conducted using Ferromagnetic Resonance (FMR), have shown a change in linewidth and the increased damping parameter in MS/Py heterostructure in comparison to bare Py films on $Si/SiO_2$ substrate, indicating spin transport from Py to MS at the interface. We observed similar results in the case of VMS/Py heterostructure too. However, the change in the linewidth and damping is smaller in VMS/Py than that in MS/Py, indicating reduced (only 15%) SOC in VMS compared to MS. Here, enhanced electrical conductivity and negligible reduction in SOC strength along with ferromagnetic ordering are demonstrated in VMS.

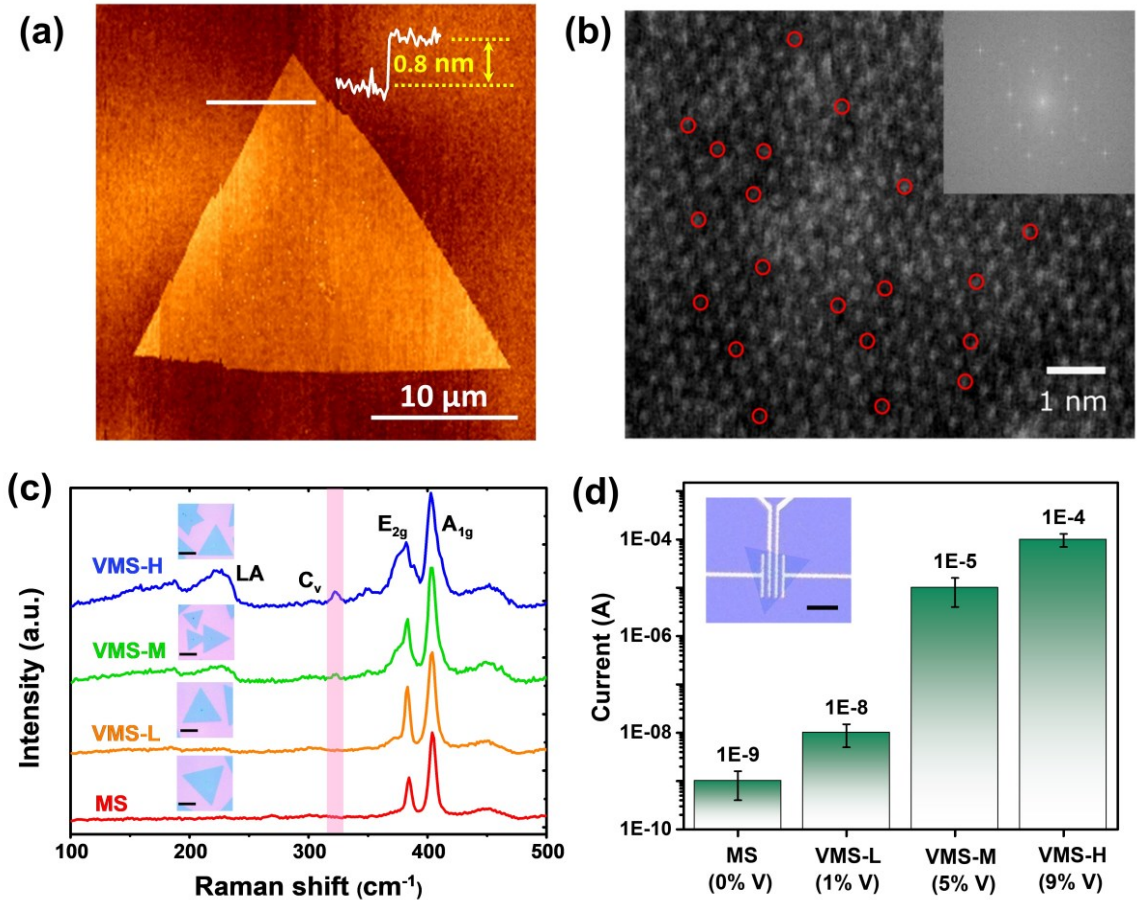

**Figure 1:** (a) AFM image and thickness profile of high VMS-H, indicating the formation of monolayer crystals of triangular shape. (b) STEM image of VMS where the marked red circle showing the presence of V in the $MoS_2$ lattice. V atom distributed randomly in the lattice of $MoS_2$. The inset shows the FFT spectrum of the image where the hexagonal distribution is clearly visible. (c) Raman spectra of MS and different VMS using 532 nm laser excitation. inset shows the optical images with a scale bar of 20 μm. The highlighted area represents the peak corresponding to V-doping, which clearly indicates that with increasing V concentration, the peak increases. (d) Bar diagram plot of current values for MS and different VMS samples. Inset is the optical image of the device with a scale bar of 30 μm. Channel length (10 μm) and $V_{DS}$ (1 V) are kept the same in all the devices, indicating the increase in conductivity with V doping.

**Results and Discussion**

The MS and different atomic percentages of V doping in MS are synthesized on a Si/SiO$_2$ (300 nm) substrates. In this study, we have optimized three different atomic percentages of V doping [VMS-L (1% V), VMS-M (5% V), and VMS-H (9% V)] using the same precursor (V$_2$O$_5$) with the addition of a small amount of salt (NaCl). The details of elemental quantification are discussed elsewhere.[22] Figure 1(a) presents the atomic force microscopy (AFM) image of the CVD-grown monolayer crystal of VMS-H. The height profile reveals that the layer thickness is ~ 0.8 nm, indicating the formation of monolayer crystals.[28] Field-emission scanning electron microscopy (FESEM) images of CVD-grown monolayer MS and various VMS are provided in the supporting information, Figure S1. Figure 1(b) displays the high-resolution transmission electron microscopic (HR-TEM) image of VMS (9%) and the fast Fourier transform (FFT) pattern (inset of Figure 1(b)) indicating crystallinity and hexagonal symmetry, confirming the hexagonal lattice structure of VMS.

Figure 1(c) depicts the Raman spectra of MS and VMS samples, with an inset showing optical images of the crystals. The V doping did not significantly affect the morphology of the crystals, as evidenced by the triangular-shaped crystal formation observed in both the optical and FESEM images for MS as well as VMS. Pristine MS exhibits the characteristic Raman modes at 384 cm$^{-1}$ and 403 cm$^{-1}$, respectively, separated by ~ 19 cm$^{-1}$, as depicted in Figure 1(c).[29,30] In VMS, in addition to these two characteristic peaks of MS, several new Raman active modes emerged between 125 cm$^{-1}$ to 270 cm$^{-1}$, termed as LA mode. The intensity of the LA mode allows for the calculation of the percentage of vanadium doping in the MoS$_2$ system, as reported in our previous work.[22] Alongside the LA mode, a new peak at 325 cm$^{-1}$ (highlighted in pink) arises due to the presence of vanadium atoms doped into the MS lattice, which is sensitive to doping concentration.[17,18] In the case of VMS, the separation between E$_{2g}$ and A$_{1g}$ peaks increases from 19 to 24 cm$^{-1}$, and the peaks also become broader with increasing

V concentration, attributed to the strain induced by V doping.[31] Figure 1(d) illustrates a bar diagram plot of current values for MS and VMS with varying V doping levels. The channel length (10 μm) and the $V_{DS}$ (1 V) are maintained the same for all the devices (details in supporting information).

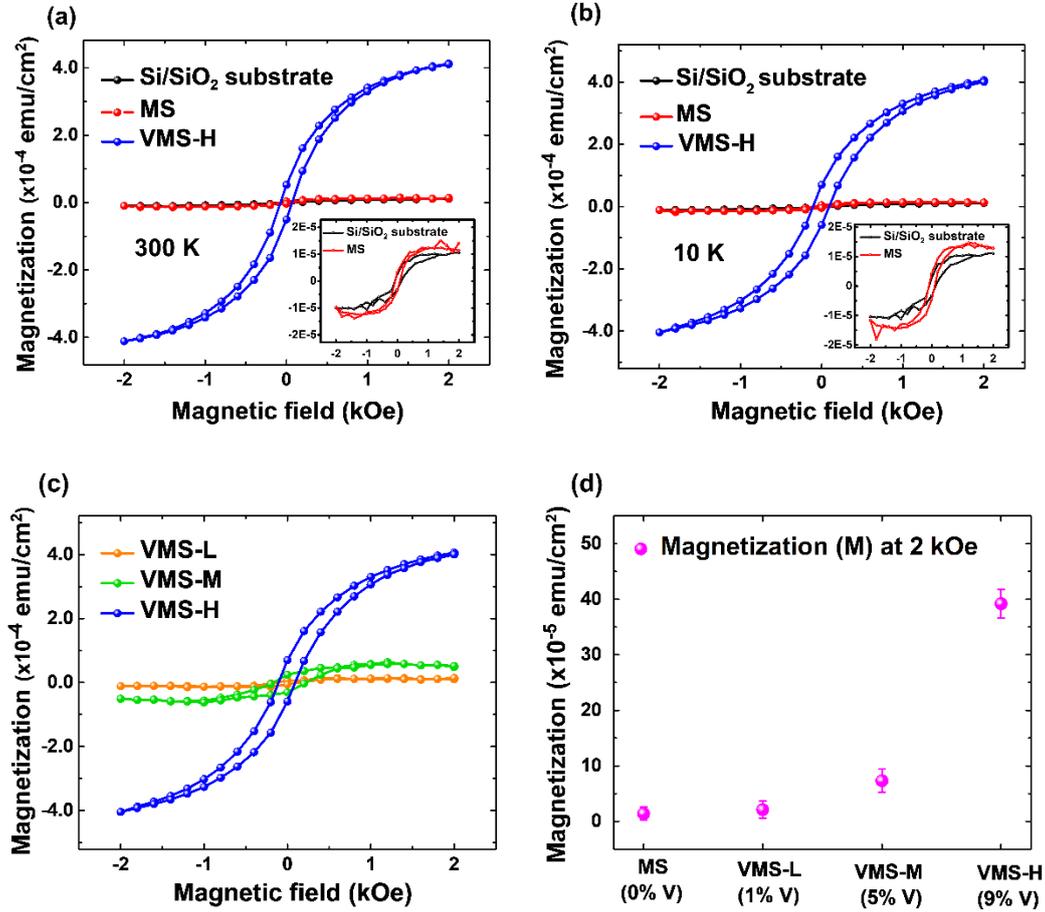

**Figure 2:** (a) The M-H curves at 300 K for Si/SiO$_2$ substrate, MS, and VMS-H at 300 K. (b) M-H curves for Si/SiO$_2$ substrate, MS, and the VMS-H at 10 K. (c) Comparison of M-H curves for the different percentages of V-doping i.e., VMS-L (1 atomic%), VMS-M (5 atomic%), and VMS-H (9 atomic%). d) Magnetization (M) at 2 kOe is compared for MS different V concentrations, indicating an increase in the magnetization with V content.

The current ($I_D$) for the MS falls within the range of $10^{-9}$ A in the measured voltage window ($V_{DS}$ =1 V), whereas for VMS, the current values are in the range of ~$10^{-8}$, $10^{-5}$, and $10^{-4}$ A for VMS-L, VMS-M, and VMS-H, respectively. The room temperature conductivity of the MS,

VMS-L, VMS-M, and VMS-H are calculated as ~0.315 S/m, ~12.5 S/m, ~9091 S/m, and ~105263 S/m, respectively, indicating the enhanced electrical conductivity from MS to VMS-H.

*Static magnetic properties of Vanadium doped MoS$_2$:*

Static magnetic properties are studied using the magnetization reversal curves obtained using a SQUID magnetometer for the bare Si/SiO$_2$ substrate, Si/SiO$_2$//MS, and Si/SiO$_2$//VMS. The M-H measurements are conducted at two different temperatures, 300 K and 10 K. Figure 2(a) and 2(b) shows the M-H loops of bare Si/SiO$_2$ substrate, pristine MS, and VMS on Si/SiO$_2$ substrate at 300 K and 10 K respectively. The M-H loops indicate that the bare Si/SiO$_2$ substrate and MS samples have negligible magnetization (~1.4x10$^{-5}$ emu/cm$^2$ at 2 kOe). Interestingly, upon doping with V, the VMS samples indicate considerable magnetization, and VMS-H shows the highest magnetization (~39.2x10$^{-5}$ emu/cm$^2$ at 2 kOe with coercivity ~120 Oe) both at room temperature (Figure 2(a)) and 10 K (Figure 2(b)). As explained before, the V-V interactions of the V or S defects can induce the ferromagnetic ordering in V-doped MoS$_2$. Figures 2(a) and 2(b) show that the magnetic ordering in VMS is robust and independent of temperature change ranging from 10 K to 300 K. Figure 2(c) shows the comparison of the M-H curves of MS with different percentages of V-doped MS, i.e., low (1%), medium (5%) and high (9%). The enhancement in the magnetization with V-content, as ~ 1.4x10$^{-5}$ emu/cm$^2$ for MS, 2.1x10$^{-5}$ emu/cm$^2$ for VMS-L, 7.3x10$^{-5}$ emu/cm$^2$ for VMS-M, and 39.2x10$^{-5}$ emu/cm$^2$ for VMS-H at 2 kOe, is shown in the Figure 2d, indicating that V doping in the MS is responsible for the induced magnetic ordering.

*Magnetization dynamics and spin transport studies on MS and VMS:*

*Ferromagnetic resonance (FMR) measurements:*

FMR based study is used to understand the tunability of the SOC in MS by V-doping through indirect means using the spin-pumping method. A schematic of the FMR setup is shown in Figure S2 (a). Pure spin current (J$_S$) can be generated and injected at the ferromagnet/non-magnet (FM/NM) interface through spin pumping. The stack prepared to study the spin pumping is shown in Figure S2 (b). The magnetization precession is generated in the ferromagnetic layer by applying an external magnetic field. An external RF field is applied, the resonance occurs when the frequency of precession matches with the slope external RF signal. At FMR, large precession leads to the transfer of angular momentum at the interface between FM and NM. The angular momentum flow is associated with pure spin current. The spin pumping phenomenon has been explained by Tserkovnyak *et al.* in terms of an increase in the Gilbert damping parameter ($\alpha$) due to the loss of spin angular momentum at the interface when spin current is injected from the FM layer to the adjacent NM layer.[32,33] As a result, the absorption spectrum gets broadened, resulting in the increase in the linewidth. Therefore, the FMR-associated linewidth broadening ($\Delta H$) and increment in damping parameter is considered to be a direct method to understand the spin transport from FM to NM.[34-36] The spin transport across the interface is closely related to the SOC of the NM layer. Therefore, the change in linewidth broadening and damping parameter can be used to understand the SOC strength of NM. The NM layers in this study are MS and VMS (VMS-H will be termed as VMS from here on).

The FMR spectra is analyzed using the following equations as shown in the literature:[36-40]

$$\frac{dI}{dH} = k_1 \cdot \left[ \frac{\left(\frac{\Delta H}{2}\right)(H-H_{res})}{\left\{(H-H_{res})^2 + \left(\frac{\Delta H}{2}\right)^2\right\}^2} \right] + k_2 \cdot \left[ \frac{\left(\frac{\Delta H}{2}\right)^2 - (H-H_{res})^2}{\left\{(H-H_{res})^2 + \left(\frac{\Delta H}{2}\right)^2\right\}^2} \right] + Offset + H.Slope \quad (1)$$

$$\Delta H = \Delta H_o + \frac{4\pi \alpha_{eff} f}{\gamma} \quad (2)$$

$$f = \frac{\gamma}{2\pi} \sqrt{(H_{res} + H_k + 4\pi M_{eff})(H_{res} + H_k)} \quad (3)$$

We have deposited 8 nm permalloy (Py, $Ni_{81}Fe_{19}$) as FM using magnetron sputtering technique on bare $Si/SiO_2$, MS on $Si/SiO_2$, and VMS on $Si/SiO_2$ samples to perform spin pumping studies using FMR to understand the changes in SOC. Figure 3(a) shows the FMR response of Py deposited on $Si/SiO_2$. The resonance field ($H_{res}$) and linewidth ($\Delta H$) are extracted by fitting the FMR spectra at each frequency (*f*) to equation (1). The fitting of the spectrum is shown in the supporting information, Figure S3 (a). Figure 4(b) shows the frequency dependence of linewidth and the damping ($\alpha$) parameter, i.e. 0.0079 is extracted using equation (2), where $\gamma$ is the gyromagnetic ratio and $\gamma/2\pi$ = 0.00282 GHz/Oe. The effective magnetization ($4\pi M_{eff}$) of the sample is extracted to be 8730 Oe by fitting the $H_{res}$ vs $f$ to the equation (3) as shown in Figure 3(c), where $H_k$ is the magnetic anisotropy field. Similar measurements and analysis have been carried out for MS/Py and VMS/Py. The parameters α, $4\pi M_{eff}$, inhomogeneous linewidth broadening (δΔH) and the change in FMR-associated linewidth broadening (ΔH) are recorded for all the three samples and are shown in Table ST2 (supporting information). The inhomogeneous linewidth broadening is observed to be small (<10 Oe) in all samples indicating the uniformity and quality of the Py films deposited. The Si/SiO$_2$//Py (8 nm) sample is used as the reference sample. We have observed an increase in α both in MS/Py (0.0098) and VMS/Py (0.0085) samples compared to the reference sample (0.0079). Figure 3(d) shows the comparison of linewidth at different frequencies for all the samples. From the figure, we can understand that there is an increment in the linewidth in MS/Py and VMS/Py compared to the reference sample. Similar results have been observed in the samples where the thickness of the Py is 14 nm. Figure S3 (b) shows the comparison of linewidth broadening in Si/SiO$_2$//Py (14 nm), Si/SiO$_2$//MS/Py (14 nm) and Si/SiO$_2$//VMS/Py (14 nm). As explained earlier, the increase in α and linewidth broadening indicates the spin transport from Py into MS and VMS. The change is linewidth broadening (δΔH) and the change in α is higher in the case of MS/Py as compared to VMS/Py. In general, higher the SOC strength of the material, higher will be the loss of spin angular momentum at the interface, thereby higher the change in α and ΔH. It is to be noted that the change in damping parameter is around 15% from MS to VMS, indicating a small drop in SOC strength.

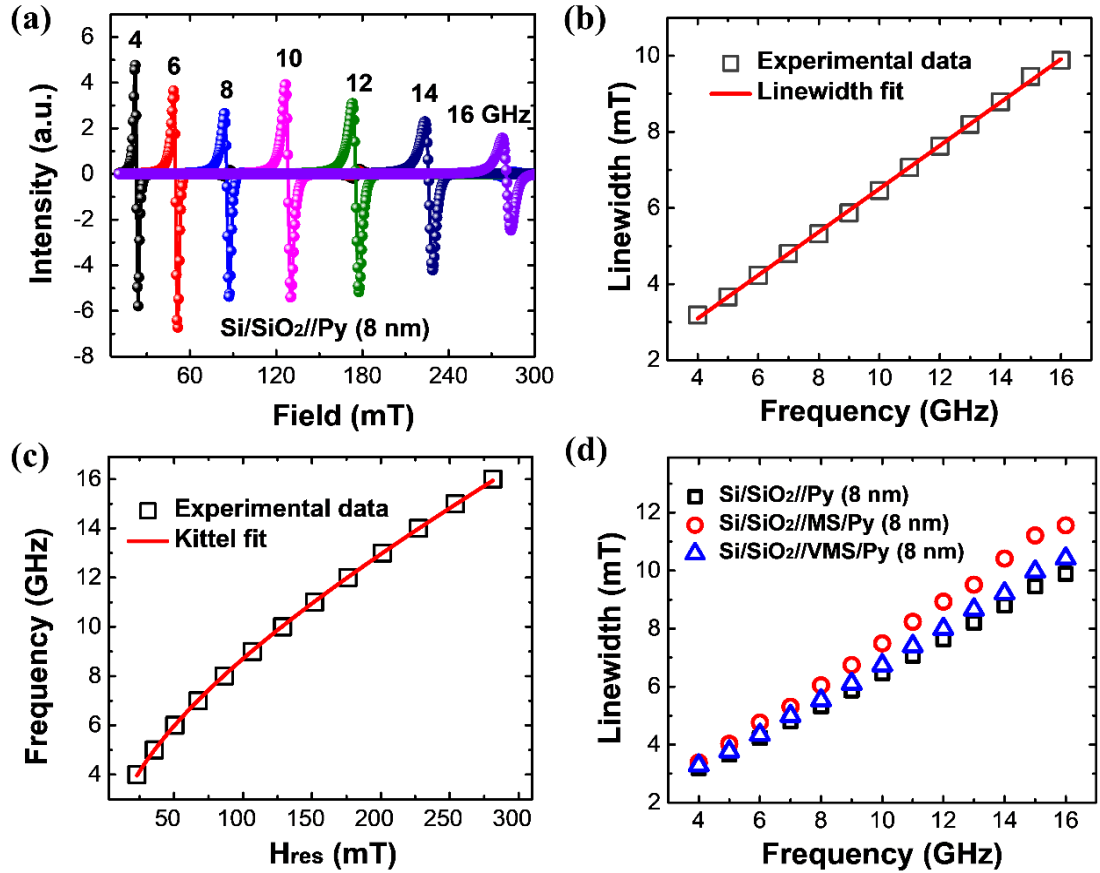

**Figure 3:** (a) FMR spectrum of $Si/SiO_2//Py$ (8 nm) from 4-16 GHz. (b) The experimental data of frequency vs linewidth and the fit to the equation (3), (c) The experimental data of $H_{res}$ Vs frequency and the Kittel (equation (2)), (d) Comparison of linewidth at frequencies from 4-16 GHz for the samples $Si/SiO_2/Py$ (8 nm), $Si/SiO_2//MS/Py$ (8 nm) and $Si/SiO_2//VMS/Py$ (8 nm).

**Spin-torque ferromagnetic resonance (ST-FMR) measurements:**

To demonstrate spin current generation in VMS and its injection into Py, we have performed ST-FMR measurements. The reprentative stack and the ST-FMR measurement schematic are shown in Figure 4 (a). In ST-FMR, the RF charge current ($J_{rf}$) is injected through a patterned co-planar waveguide (CPW). This charge current generates a spin current in VMS, which then travels towards Py and exerts a torque on the magnetization (M) of Py, which is precessing around the effective magnetic field. This results in the change of anisotropic magnetoresistance, which peaks at the ferromagnetic resonance condition (where the RF current frequency and the precessional frequency of magnetization matches).

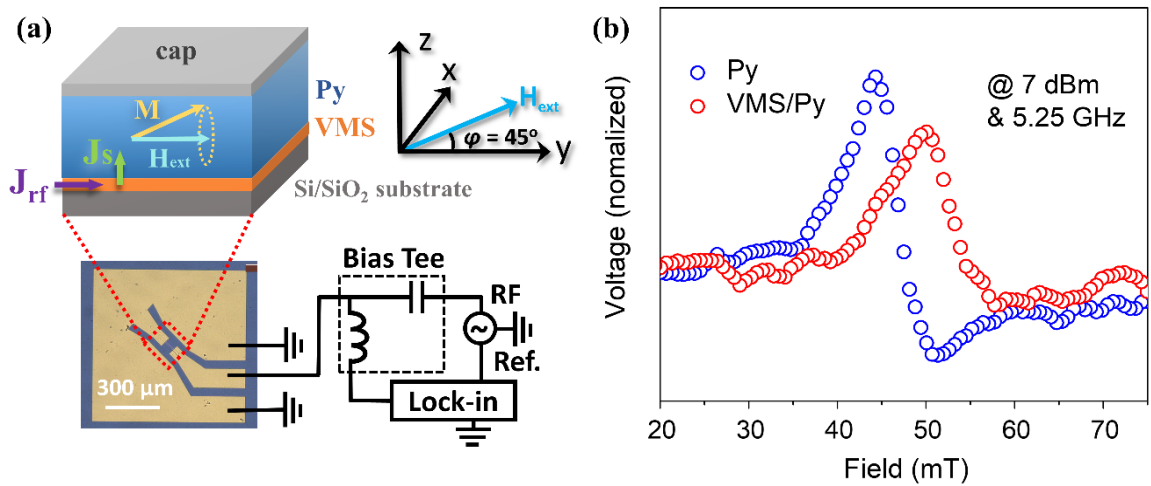

**Figure 4:** (a) Schematic representation of the phenomenon involved in ST-FMR measurements. Coplanar wave guide devices (optical image) and the measurement schematic are shown in the inset. (b) ST-FMR response of $Si/SiO_2//Py$ (8 nm) and $Si/SiO_2//VMS/Py$ (8 nm) samples.

ST-FMR measurements are performed on the patterned microwires of Py (8 nm) and VMS/Py (8 nm) samples on $Si/SiO_2$ substrate (details are provided in the Experimental methods). Here, $Si/SiO_2//Py$ (8 nm) sample is considered as the reference sample to demonstrate the spin transport in $Si/SiO_2//VMS/Py$ (8 nm). The ST-FMR spectra of Py and VMS/Py at 5.25 GHz are shown in Figure 4(b). The data is fit to the Lorentzian form using equation (4) to extract the linewidth (ΔH), resonance field ($H_{res}$), and symmetric (S) and antisymmetric (A) components of the voltage signal ($V_{mix}$). [8,41]

$$V_{mix} = S \frac{\Delta H^2}{\Delta H^2 + (H_{ext} - H_{res})^2} + A \frac{\Delta H(H_{ext} - H_{res})}{\Delta H^2 + (H_{ext} - H_{res})^2} \quad (4)$$

It is observed that the $H_{res}$ and the ΔH are increased in VMS/Py compared to the Py, as shown in Figure 4(b). In general, the spin torque exerted on the magnetization of Py leads to this change in $H_{res}$ and ΔH.[41] The spin Hall efficiency ($\xi_{ST-FMR}$) of a material is estimated using the equation (5), where $M_s$ is the saturation magnetization, $t_{NM}$, and $t_{FM}$ are the thickness of the non-magnet (spin Hall layer), and the ferromagnet, respectively.

$$\xi_{ST-FMR} = \frac{S}{A} \frac{e \mu_o M_s t_{NM} t_{FM}}{\hbar} \sqrt{1 + \frac{4\pi M_{eff}}{H_o}} \quad (5)$$

For a 2D material like VMS, the origin of spin-orbit torques mainly stems from the interfacial Rashba-Edelstein effect. In such a scenario, estimating spin Hall efficiency using equation (5) becomes challenging due to the interfacial origin of SOC.[5,42] However, the S and A parameters directly represent damping-like and field-like torques, and are the by-products of the spin-orbit torque from VMS. This suggests that the S/A ratio can be considered to understand the change in spin Hall efficiency qualitatively. It is shown in equation (5) that the S/A ratio is directly proportional to the charge-to-spin conversion efficiency. From the extracted S and A values of Py and VMS/Py at 5.25 GHz, we estimated the S/A ratio of both the samples.

The S/A ratio, a measure of spin Hall conductivity, of Si/SiO$_2$//VMS/Py (8 nm) is found to be ~1.44, better than that of Si/SiO$_2$// Py(8 nm) ~0.72 (Table ST3). This increment in the S/A ratio from Py to VMS/Py indicates the spin torque effect on magnetization of Py due to the spin current from VMS to Py. Moreover, the S/A ratio values of platinum, one of the widely established spin-Hall materials, of thickness ~15 nm and 6 nm are reported as ~0.08 and 0.63, respectively.[41]

The photoluminescence-based studies presented in the supporting information (Figure S4 and Table ST1) indicate that the SOC of MS tends to gradually reduce with doping percentage of V (as shown in the A and B exciton emission as a measure of the SOC strength). This supports the aforementioned discussion that the change in SOC of MS after V substitutional doping leads to the change in damping parameter and the linewidth broadening. To further support the relatively high SOC in MS and VMS and its tunability, DFT based studies are conducted.

### *DFT Analyses*:

To investigate the effects of vanadium doping in MoS$_2$, DFT calculations are performed on a 2x2 MoS$_2$ cell with one Mo atom replaced by a vanadium atom. The complete description of the DFT formulation is given in the supporting information. As discussed in our previous work,[17] the valence band of MoS$_2$ splits by ~150 meV and the conduction band by ~10 meV due to SOC. Introducing vanadium disturbs the lattice and alters the electronic structure, making VMS more metallic. V-doping induces spin-splitting at K and K' points, leading to spin polarization (Figure S5a). The SOC-based DFT calculations reveal that SOC causes unequal splitting at these valleys, lifting degeneracy and causing a ~145 meV energy shift (Figure S5b), which enhances spin polarization in VMS. These results indicate that V-doping in MoS$_2$ induces magnetic behavior through SOC modification and spin polarization.

In conclusion, achieving high spin-orbit coupling (SOC) with large electrical conductivity is one of the bottlenecks in the development of spin Hall materials. Here, we demonstrate the potential of vanadium (V) doping in MoS$_2$ monolayers, leading to high electrical conductivity (~10$^5$ S/m) and SOC strength 2D layers having room temperature magnetic ordering. Degenerate level of V substitutional doping (9 atomic%) in MoS$_2$ is shown to be converting a low conductivity MoS$_2$ to a metal-like one having large magnetization 39.2x10$^{-5}$ emu/cm$^2$ at 2 kOe. A systematic enhancement in the magnetization is observed with V doping and the origin of spin polarization and SOC modification are verified using DFT based calculations and photoluminescence-based studies. Magnetization dynamics studies using FMR indicate a change in linewidth and the increased damping parameter in MS/Py heterostructure in comparison to bare Py films on Si/SiO$_2$ substrate, demonstrating spin transport from Py to MS. We observed similar results in the case of VMS-Py (VMS/Py) heterostructure too. However, the change in the linewidth and damping are smaller than that in MS/Py, indicating smaller SOC in VMS as compared to MS. ST-FMR measurements demonstrated the charge to spin conversion in VMS. The ratio of symmetric and the antisymmetric components of the ST-FMR voltage signal is found to be higher in VMS/Py (1.44) stack as compared to Py (0.62), indicating the spin injection from VMS to Py. The enhanced electrical conductivity with modified SOC strength and magnetic ordering shown in VMS highlight the potential of VMS as an efficient 2D spin-orbit torque material for spintronic applications.

**Acknowledgements:**

We thank Prof. Subhankar Bedanta (NISER Bhubaneswar, India), and Abhisek Mishra (NISER Bhubaneswar, India) for their fruitful discussions on this project. We thank Mr. Munawar M and Mr. Rohiteswar Mondal (IIT Hyderabad, India) for assistance in microfabrication and few of the magnetic measurements, respectively. We thank Prof. Raul Arenal (INMA, Spain) and Mr. Sreekant Anil for STEM and SEM measurements. Authors from TIFR acknowledge the Department of Atomic Energy, Government of India, for the funding support (project identification number: RTI4007). NN acknowledges SERB-SUPRA grant for support under Grant Number SPR/2020/000220 entitled 'Development of Spin-correlation Engineered Transition Metal Dichalcogenides and Novel Light-assisted Methods for probing them.

**Associated Content:**

The supporting information contains experimental methods, SEM images, schematic of FMR setup, fitting of FMR spectra and corresponding Tables, Photoluminesence spectra and table corresponding to their peak position, and DFT analysis.

**Author Contribution:**

TNN in discussion with CM conceived the idea. KRS, MT, AH, CM, and TNN conceptualized, designed, analyzed, and concluded the project. KRS and MT carried out the major part of the experiments. DM and SM performed magnetization measurements. DM carried out the electrical conductivity measurements. MSD helped to conduct the ST-FMR measurements. AL performed the DFT-based calculations. All the authors contributed to the discussions and paper writing.

# Supporting Information

**Vanadium Doped Magnetic MoS$_2$ Monolayers of Improved Electrical Conductivity as Spin-Orbit Torque Layer**

Krishna Rani Sahoo,[1,2#] Manoj Talluri,[3#] Dipak Maity,[1#] Suman Mundlia,[1] Ashique Lal,[1] M. S. Devapriya,[4] Arabinda Haldar,[4] Chandrasekhar Murapaka,[3]* and Tharangattu N. Narayanan,[1]*

[1]*Materials & Interface Engineering Laboratory, Tata Institute of Fundamental Research Hyderabad, Serilingampally Mandal, Hyderabad 500046, India.*

[2]*Institute of Physics, University of Münster, Wilhelm-Klemm-Str. 10, 48149 Münster, Germany*

[3]*Department of Materials Science and Metallurgical Engineering, Indian Institute of Technology-Hyderabad, 502284, Telangana, India.*

[4]*Department of Physics, Indian Institute of Technology-Hyderabad, 502284, Telangana, India.*

**Experimental methods:**

*Synthesis of MS and VMS*:

MS and VMS are grown using a CVD-based approach. In brief, for the MS growth, sulfur powder and $MoO_3$ powder are placed in a two-zone furnace where they function as sulfur and Mo sources for the $MoS_2$ crystal growth. Sulfur is kept at 200 °C while $MoO_3$ is placed in a high-temperature zone kept at 700 °C. The $Si/SiO_2$ substrate is kept over an alumina boat containing $MoO_3$ powder. A similar method is followed for V-doping but there, instead of only $MoO_3$, a mixture of $MoO_3$ and $V_2O_5$ is used during the growth. A small amount of NaCl is also mixed with this mixture to help the decomposition of the metal at comparatively lower temperatures. All the experiments are carried out in nitrogen gas atmosphere where the nitrogen flow is kept at 200 sccm. To prevent multilayer growth, the furnace is suddenly cooled down to room temperature after the growth time of 15 minutes.

*Characterizations*:

Optical images are captured using a microscope, while Raman and PL spectra are carried out using a Renishaw Invia micro-Raman spectrometer with a 532 nm excitation (2.33 eV). The sample morphology is examined using scanning electron microscopy (FESEM, JEOL JSM-7500F). Atomic force microscopy measurements are conducted in dynamic mode and analysed using Gwyddion software. Scanning transmission electron microscopy (STEM) imaging is performed using a probe aberration-corrected Thermo Fisher Scientific Titan TEM microscope (Low Base) operated at 100 keV.

Four probe devices are fabricated using laser lithography, and metal (Cr/Au) deposition of (6/55 nm) is achieved through thermal evaporation. All the current-voltage measurements are carried out in an ambient environment using a Keithley 2450 source meter.

*Magnetization (M-H) measurements*:

Magnetic measurements are performed using a Cryogenic Ltd. S700X magnetometer equipped with a 70 kOe magnet.

*Deposition of Py thin-film:*

DC magnetron sputtering is employed to deposit Permalloy ($Ni_{81}Fe_{19}$ or Py) thin film on $MoS_2$ (MS/Py) and V-$MoS_2$ (VMS/Py). The base pressure of the sputtering chamber is maintained at $5 \times 10^{-7}$ mbar for the depositions. The films are grown at a deposition pressure of $5 \times 10^{-3}$ mbar and at a deposition rate of 0.13 nm/s.

*FMR measurements*:

Ferromagnetic resonance (FMR) is used to study the spin transport across non-magnetic (NM) materials $MoS_2$ (MS) and V-doped $MoS_2$ (VMS) and Py (FM). In FMR, the sample is placed in a flip-chip configuration on a coplanar waveguide that carries RF current ranging from 4-16 GHz from the RF generator. The RF magnetic field generated between the S and G lines excites the magnetization of the permalloy. An electromagnet is placed in an in-plane configuration for the sample with a magnetic field range of -3 kOe to 3 kOe. The external magnetic field is modulated by Helmholtz coils that are attached to the electromagnet's poles at a modulation frequency of 490 Hz. This modulation frequency serves as the reference for the lock-in amplifier to detect the FMR signal. Due to the microwave excitation by the RF field, the FM layer will be driven into precession about the effective magnetic field. Resonance occurs when the precession frequency of the magnetization of FM matches with the frequency of the RF field. At resonance conditions, the sample absorbs RF energy, which is detected by an RF diode that feeds the signal to the lock-in amplifier. The lock-in-amplifier then enhances the signal-to-noise ratio of the signal. The derivative of the absorption intensity (dI/dH) is plotted against the external magnetic field H.

*CPW-device fabrication:* Coplanar waveguide devices for ST-FMR measurements are prepared using conventional photolithography techniques using maskless and lift-off methods. Wires of the samples with dimensions 100 x 10 μm are prepared. In the second step, the contact pads of Ta (5 nm)/Au (80 nm) are sputter deposited.

*ST-FMR measurements:* Spin torque ferromagnetic measurements are employed to understand the spin transport from VMS to Py. An electromagnet with a DC power supply is used to sweep a DC magnetic field from 0 to 100 mT. RF current with 7 dBm power and frequency ranging from 4.5 GHz to 6.25 GHz is fed to the sample from an RF generator using a G-S-G probe. The change in resistance due to the anisotropic magnetoresistance(AMR) is recorded in the form of voltage using the same probe. A bias-Tee is used to feed RF current and measure DC voltage using the same probe. RF input of the bias-Tee is connected to the RF generator; RF output is connected to the G-S-G probe, and DC output coming from G-S-G is connected to a nano voltmeter to measure DC voltage. The magnetic field is applied at an angle of 45 deg. to the sample to enhance the change in resistance at the resonance condition.

*DFT Calculations*:

First principle DFT-based calculations of V-doped $MoS_2$ are carried out using QUANTUM ESPRESSO 6.8 package with a plane-wave basis set (50-Ry cutoff) and PAW (Projector-Augmented Wave) pseudopotentials. The exchange-correlation energy of electrons is approximated using the Generalized Gradient Approximation with Hubbard Potential (GGA + U). All the calculations are spin-polarized, considering the spin-orbit coupling (SOC) effect with non-collinear calculations, as implemented in QUANTUM ESPRESSO. Based on the previous reports and findings, the Hubbard U factor 3 eV is adopted for the vanadium element. A vacuum of ~15 Å is added, separating adjacent periodic images along the z-direction. The lattice parameter of the monolayer pristine $MoS_2$ unit cell is optimized to 3.17 Å using Brillouin zone sampling of $12 \times 12 \times 1$. Further electronic structure calculations are done using a $2 \times 2$ periodic supercell with and without a single vanadium dopant, where integrations over the Brillouin zone are sampled with uniform 4 × 4 × 1.

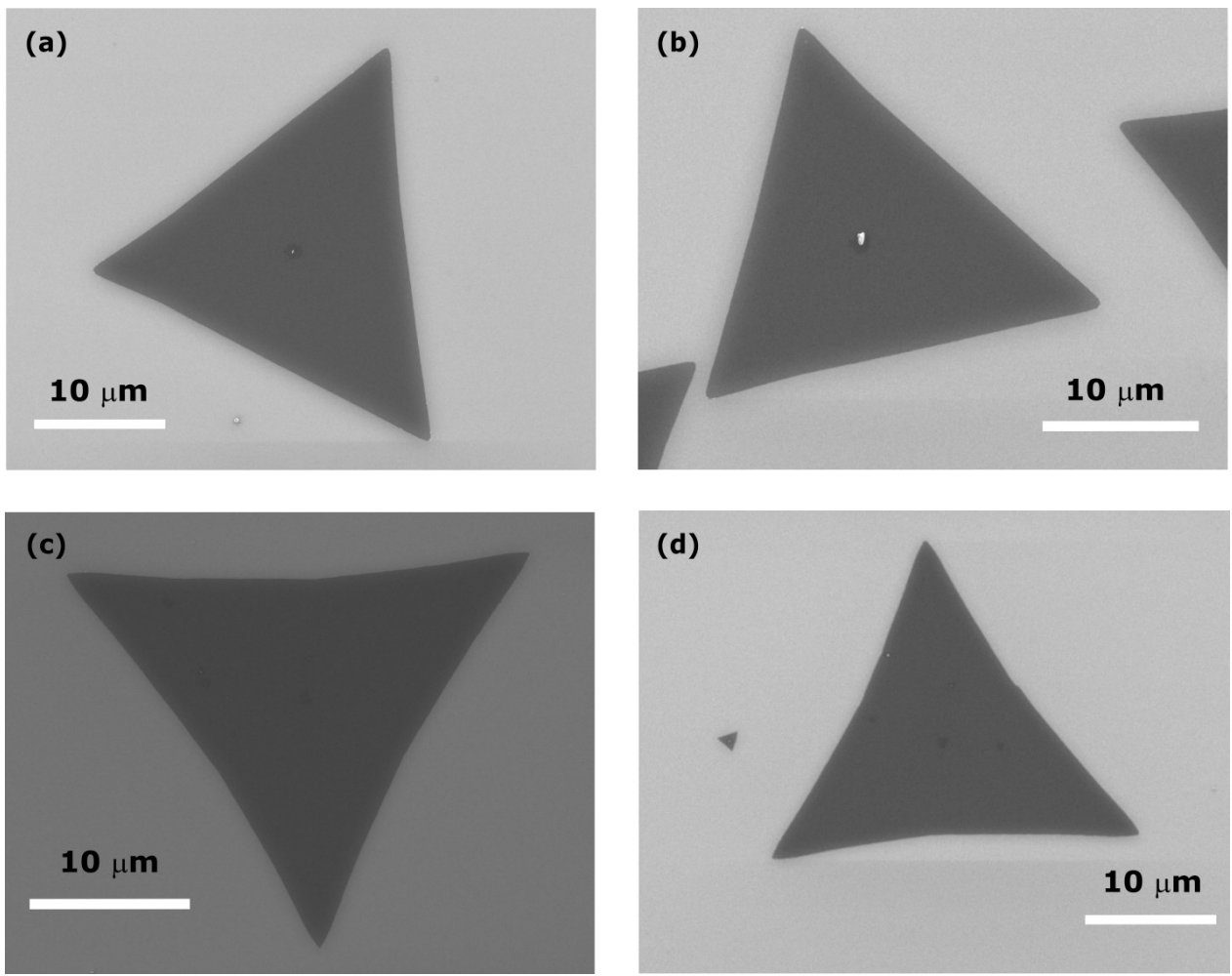

**Figure S1:** FESEM images of CVD grown monolayer a) MS b) VMS-L c) VMS-M d) VMS-H.

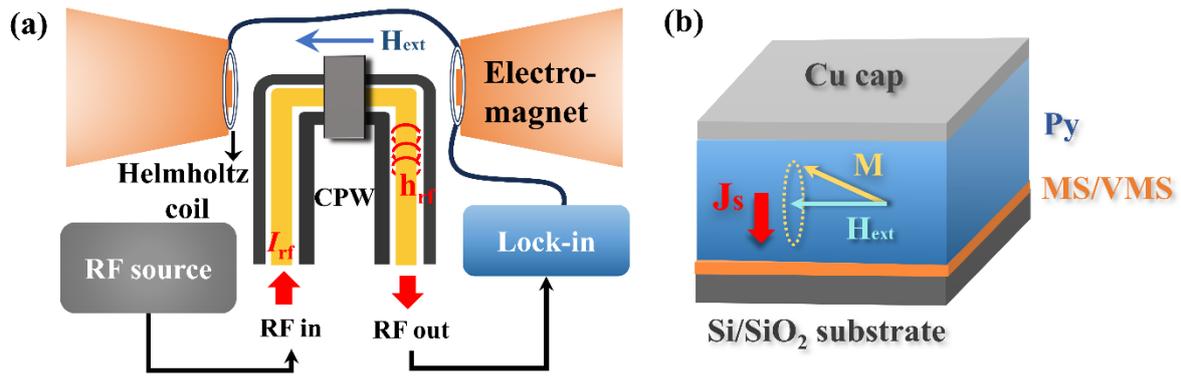

**Figure S2:** (a) Schematic representation of the FMR setup, where $I_{rf}$ is RF current provided using the RF source, $h_{rf}$ is rf field. A lock-in (Lock-in) amplifier is shown. The dark box in schematic shows the sample stack where its structure is shown in (b). (b) Schematic of the stack of a Si/SiO$_2$//MS(VMS)/Py(x)/Cu capping (x = 8, 14 nm) sample showing the magnetization precession and the spin current flow during spin pumping, where $J_s$ is the spin current density (arrow shows the spin current direction), M is the magnetization, $H_{ext}$ is the external magnetic field.

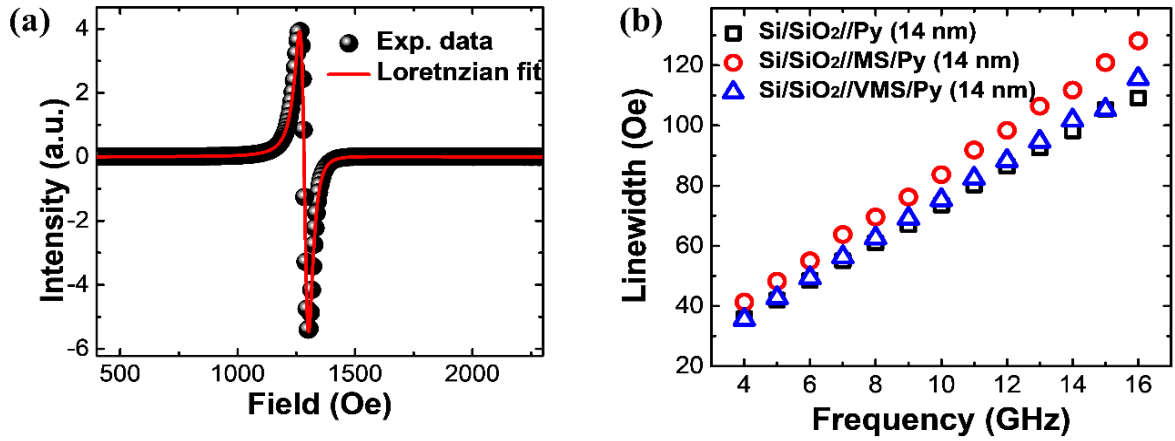

**Figure S3:** (a) FMR spectra of Si/SiO$_2$//Py (8 nm) and the Lorentzian fit to extract linewidth and resonance field, (b) Comparison of linewidth at frequencies from 4-16 GHz for the samples Si/SiO$_2$/Py (14 nm), Si/SiO$_2$//MS/Py (14 nm) and Si/SiO$_2$//VMS/Py (14 nm).

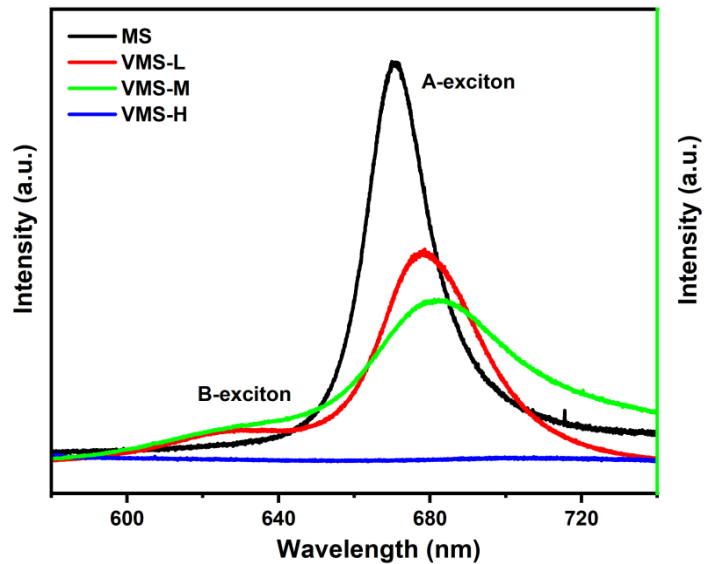

**Figure S4:** Photoluminescence (PL) data of MS, VMS-L, VMS-M, and VMS-H under 2.33 eV Laser excitation. The value of SOC is calculated from the energy difference between A-exciton and B-exciton from the PL graph and the extracted values are shown in table below.

**Table ST1**: The positions of A and B excitons and the calculated SOC strength.

| Sample | Position of A-exciton (nm) | Position of B-exciton (nm) | SOC (meV) |
|---|---|---|---|
| MS | 669 | 618 | **152** |
| VMS-L | 678 | 629 | **143** |
| VMS-M | 681 | 633 | **138** |
| VMS-H | - | - | - |

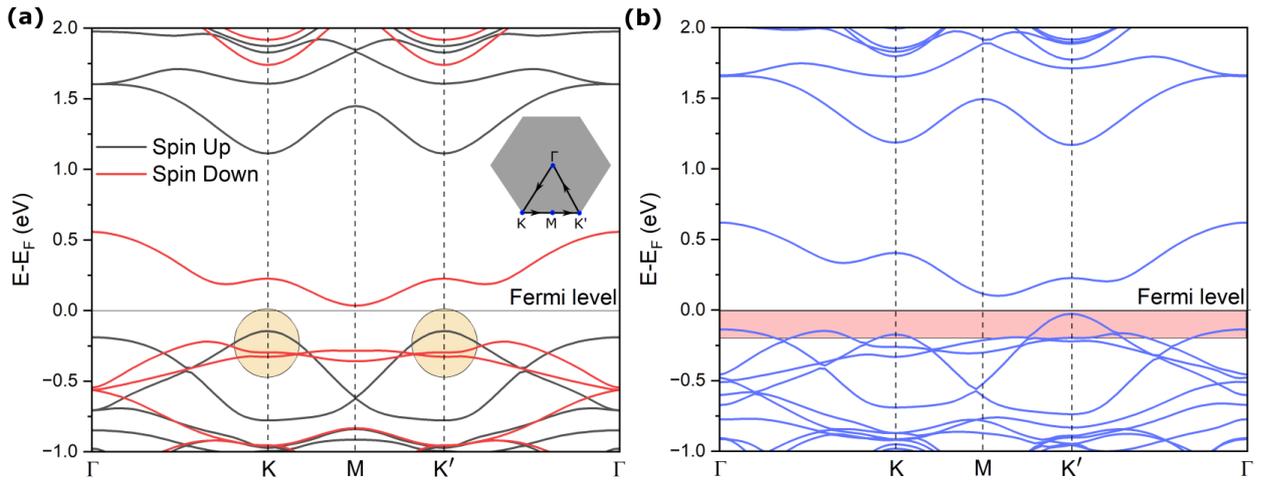

**Figure S5**: DFT frame-work showing the SOC changes with V-doping in $MoS_2$. (a) Spin resolved electronic band structure of V-doped $MoS_2$ without SOC interaction. The inset image shows the high symmetry points ($\Gamma$, K, M, K') in the surface of the Brillouin zone. The difference in spin up and spin down at K and K' are highlighted in the yellow circle indicating spin polarization. (b) Electronic band structure of V-doped $MoS_2$ with SOC interaction. The highlighted region in the plot indicates the degeneracy lifting in two valleys (K and K'). The conduction and valence band splitting at K and K' are different due to SOC change by V-doping.

**Table ST2:** Parameters extracted for the $Si/SiO_2$// Py (8 nm), $Si/SiO_2$//MS/Py (8 nm) and $Si/SiO_2$//VMS/Py (8 nm) samples from FMR.

| Sample | Damping parameter $\alpha$ | Effective magnetization $4\pi M_{eff}$ (mT) | Inhomogeneous Linewidth $\Delta H_o$ (mT) | Linewidth broadening $\Delta H$ (mT) |
|---|---|---|---|---|
| Si/SiO$_2$// Py (8 nm) | 0.0079 ± 5E-5 | 873.2 ±3.1 | 0.82 ±0.04 | 5.87 ±0.05 |
| Si/SiO$_2$//MS/Py (8 nm) | 0.0098 ± 3.5E-5 | 859.5 ±2.7 | 0.5 ±0.07 | 6.74 ±0.04 |
| Si/SiO$_2$//VMS/Py (8 nm) | 0.0085 ± 7E-5 | 868.1 ±4.2 | 0.73 ±0.03 | 6.09 ±0.02 |

**Table ST3:** Parameters extracted for the Si/SiO$_2$//Py (8 nm) and Si/SiO$_2$//VMS/Py (8 nm) samples from ST-FMR measurements.

| Sample | Linewidth, $\Delta H$ (mT) | Resonance field, $H_{res}$ (mT) | Symmetric component, S (µV) | Antisymmetric component, A (µV) | S/A |
|---|---|---|---|---|---|
| Si/SiO$_2$// Py (8 nm) | 2.65 | 46 | 2.54 | -3.54 | -0.72 |
| Si/SiO$_2$//VMS/Py (8 nm) | 3.79 | 50.9 | 2.37 | -1.64 | -1.44 |